\documentclass[structabstract]{aa}  
\usepackage{graphicx}
\usepackage{txfonts}
\usepackage{natbib}

\begin{document}
   \title{Limits in astrometric accuracy induced by surface brightness asymmetries in red supergiant stars}

    \author{E. Pasquato
          \inst{}
          \and
	 D. Pourbaix\inst{}
	 \and
	 A. Jorissen\inst{}    
          }
   \offprints{D. Pourbaix}

   \institute{Institut d'Astronomie et d'Astrophysique, Universit\'e Libre de Bruxelles, CP. 226, Boulevard du Triomphe, B-1050 Bruxelles, Belgium\\
              \email{}
       }

   \date{Received ; accepted }

  \abstract
   {}
   {Surface brightness asymmetries are a very common feature of stars. Among other effects they cause a difference between the projected barycentre and photocentre. The evolution of those surface features makes this difference time-dependent. In some cases, e. g. for supergiant stars, the displacement can be a non-negligible fraction of the star radius $R$, and if $R>1$ AU, of the parallax.}
   {We investigate the impact of surface brightness asymmetries on both the Gaia astrometric solution and the data processing flow with a theoretical approach.}
   {We show that when the amplitude of the displacement is comparable to the epoch astrometric precision, the resulting astrometric solution of a genuine single star may be, in some cases, of low quality (with some parameters up to 10$\sigma$ off). In this case, we provide an analytical prediction of the impact of the photocentre motion on both $\chi^2$ and the uncertainty in the astrometric parameters. Non-single star solutions are found, if allowed for the closest stars. A closer look at the parameters of the orbital solutions reveals however that they are spurious (since the semi-major axis is smaller than either its error or the stellar radius). It is thus possible to filter out those spurious orbital solutions. Interestingly, for the stocastic solutions, the stochastic noise appears to be a good estimate of the photocentric noise.}{} 
 

   \maketitle
%

\section{Introduction}
The Gaia mission \citep{perry2001,lind2008} will lead to the determination of the astrometric parameters of one billion objects with unprecedented precision. During its five years of satellite operations, every detected star will be observed on average 80 times, thus yielding its position ($\alpha_0$, $\delta_0$) at the reference epoch, its parallax ($\varpi$), and its proper motion ($\mu_{\alpha^*}=\mu_{\alpha} \cos \delta$, $\mu_{\delta}$). Additional parameters, such as the size of the orbit in the case of a binary, can also be adjusted. All these parameters derive from a $\chi^2$ fit. To allow the comparison between models (with different number of parameters) the $F2$ statistics will be evaluated \citep{wilson1931}:
\begin{equation}
F2=\sqrt{\frac{9\nu}{2}}\left[\sqrt[3]{\frac{\chi^2}{\nu}}+\frac{2}{9\nu}-1\right],
\label{eq:GoF}
\end{equation}
where $\nu$ is the number of degrees of freedom (the number of measurements minus the number of parameters of the fit). If $\chi^2$ does follow a chi-square distribution with $\nu$ degrees of freedom, then $F2$ follows a $N(0,1)-$distribution irrespective of $\nu$. A fit with a $F2$ larger than, say, 3 is considered poor.\\
Phenomena such as large convective cells or magnetic spots cause time-varying brightness asymmetries on the surface of stars. Their presence induces a time-dependent displacement of the photocentre with respect to the barycentre that possibly alters the astrometric parameters derived by Gaia \citep{bastian2005, eriksson2007}.
\cite{2005svensson} studied the dependence of the photocentric variability on the stellar parameters across the Hertzsprung-Russell diagram, from white dwarfs to red giants, using CO$^5$BOLD hydrodynamical models. They found a linear relation between the standard deviation in the photocentre displacement ($\sigma_P$) and the pressure scale height at optical depth unity.\\
There is only one red-supergiant hydrodynamical 3D model available in literature, \citep{chiava2009}, that predicts a standard deviation for the photocentre motion driven by convection of about 0.1 AU. This value, however, might be underestimated because of the approximations intrinsic to the model. \cite{chiava2010} extended the analysis of \cite{2005svensson} with this red supergiant model. They argue that a radical change in the convective pattern is expected between the giant and supergiant stages but the large gap in $T_{\rm eff}/g$ between the two available models does not allow a law for $\sigma_P$ as a function of $T_{\rm eff}/g$ to be unequivocally constrained. Hence, they proposed two different laws and performed a first analysis of the impact of this photocentric noise on the parallax, estimating the number of supergiants and giants for which it may be relevant.\\
The aim of this work is to investigate and provide an analytical prediction of the effects of these time-varying surface brightness asymmetries on the astrometric parameters derived by Gaia and to study their impact on the Gaia astrometric data processing. On the basis of the 3D supergiant model from \cite{chiava2009}, we simulated the photocentre motion due to convective cells and added it to Gaia simulated data (Sect.~2). We discuss the results of the astrometric fit of those data. In Sect.~3, we restrict our study to the single-star astrometric fit. Only five parameters ($\alpha_0$, $\delta_0$, $\varpi$, $\mu_{\alpha^*}$, $\mu_{\delta}$) are derived in this solution. In Sect.~4, more complex, non single-star solutions are studied.
\section{The model}
\begin{figure}[!h]
  \includegraphics[angle=270,width=\hsize]{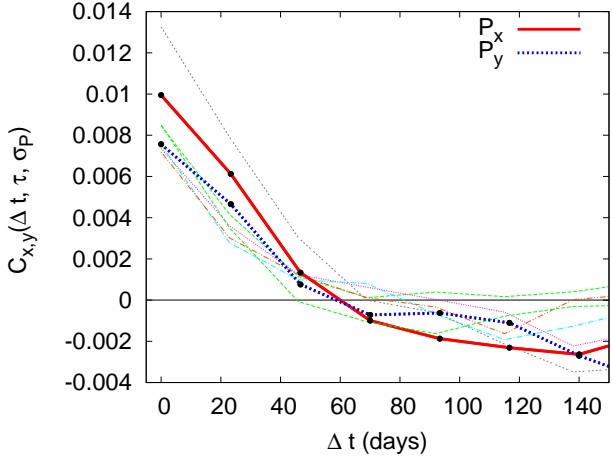}
    \caption{Thick lines: correlation function of the projections of the photocentre positions extracted from the 3D model. Thin lines: correlation function of simulated photocentre positions generated with an exponentially decreasing correlation function (see Eq.~\ref{eq:corrfuncMc}) with $\sigma_P \simeq 0.09$ AU and $\tau \simeq 2-4 \times 10^6$s sampled every 23 days.}
   \label{corrfunc}
\end{figure}

Very large displacements of the photocentre are found in red supergiants, which are caused by the presence of extremely large convective cells. The flow of hot material towards the surface and the inflow of cold material around large convective cells produce huge surface brightness asymmetries. Evidence of this phenomenon are provided by interferometric observations of supergiants shch as $\alpha$ Ori \citep{bush1990,wil1992,tuth1997,youn2000,tata2007,hau2009,2011A&A...529A.163O}.
\begin{figure}[!h]
  \includegraphics[width=\hsize]{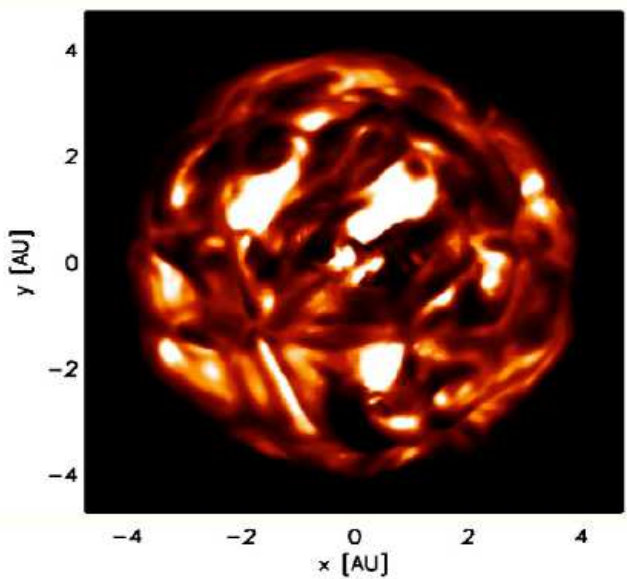}
    \caption{Snapshot of the 3D supergiant model in the Gaia $G$ band from \cite{chiava2009}. The colour scale denotes different temperatures. The 12 $M_{\odot}$ star is modelled with an equidistant numerical mesh of 235$^3$ grid points and a resolution of 8.6~$R_{\odot}$ (or 0.040 AU). The luminosity average over spherical shells and over time (i.e., over 5 years) is $L=93\,000\pm1300$~$L_{\odot}$, $T_{\rm{eff}}=3490\pm$13~K, $R=832\pm0.7$~$R_{\odot}$, and $\log g=-0.337\pm0.001$.}
   \label{snapshot}
\end{figure}
A numerical 3D model of a red supergiant was developed by \cite{chiava2009} using CO$^5$BOLD \citep{frey2002,frey2008}. 
This model was used to predict spectra and intensity maps in the Gaia $G$ band for the whole simulation time sequence, namely $\approx5$ years with time steps $\approx23$ days apart. A snapshot of the star is shown in Fig.~\ref{snapshot}.\\
The projections $P_x$ and $P_y$ of the photocentre position on either axis are central to the present study. We assume that $P_x$ and $P_y$ are random variables (of zero mean) that are both Gaussian and Markovian\footnote{A discrete process is Markovian when its future depends only on its present state and not on its past.}. Therefore the process correlation function 
\begin{equation}
C_{x,y}(\Delta t)  =   E\left[(P_{x,y}(t)-E[P_{x,y}])(P_{x,y}(t+\Delta t)-E[P_{x,y}])\right]
\label{eq:corrfunc}
\end{equation}
(where the operator $E\left[\;\right]$ denotes the mean value over $t$) reduces to \citep{doob1942}
\begin{equation}
C(\Delta t, \tau)= \sigma_{P}^2 \exp{[-\Delta t/\tau]}.
\label{eq:corrfuncMc}
\end{equation}
The random variables $P_{x,y}$ at time $t_i$ are given by
\begin{equation}
P_{x,y}(t_i)=\exp (-\Delta t_i/\tau) P_{x,y}(t_{i-1})+g_i
\label{eq:markovchain}
\end{equation}
where $\Delta t_i = t_i-t_{i-1}$, $g_i$ is extracted from a Gaussian distribution $N(0,s)$ with $s^2=\left[1-\exp (-2\Delta t/\tau)\right]\sigma_{P}^2$. Thus, all statistical properties of such a Markov chain are determined by the three parameters of its mean $E\left[P_{x,y}\right]$ (which we here assume to be zero), its variance  $\sigma^2_{P}$, and its relaxation time $\tau$, i.~e. the time after which the correlation function becomes ``negligibly small'', which in the present case is $\sigma^2_{P}/e$.
\begin{figure}[!h]
   \includegraphics[angle=270,width=\hsize]{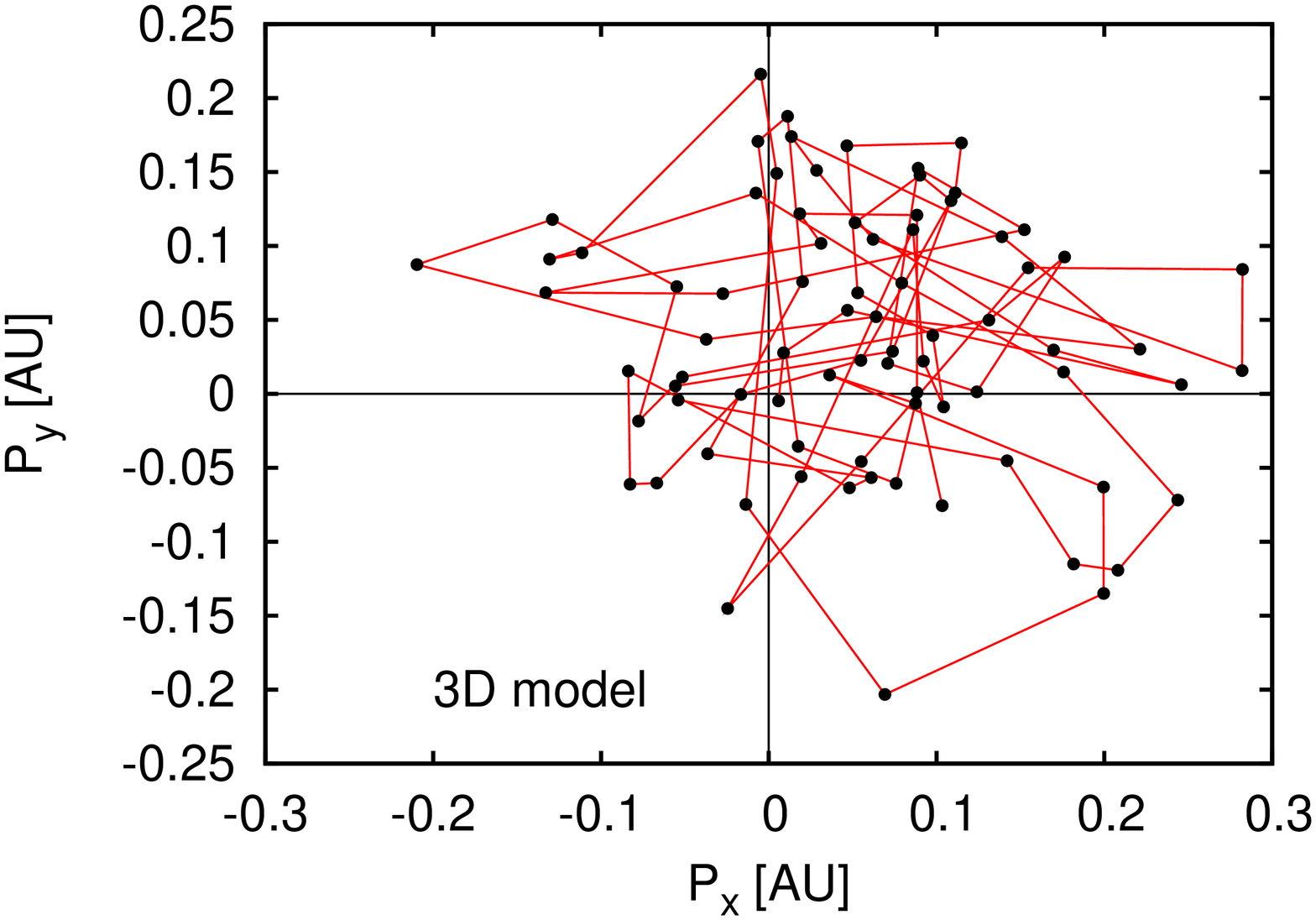}
   \includegraphics[angle=270,width=\hsize]{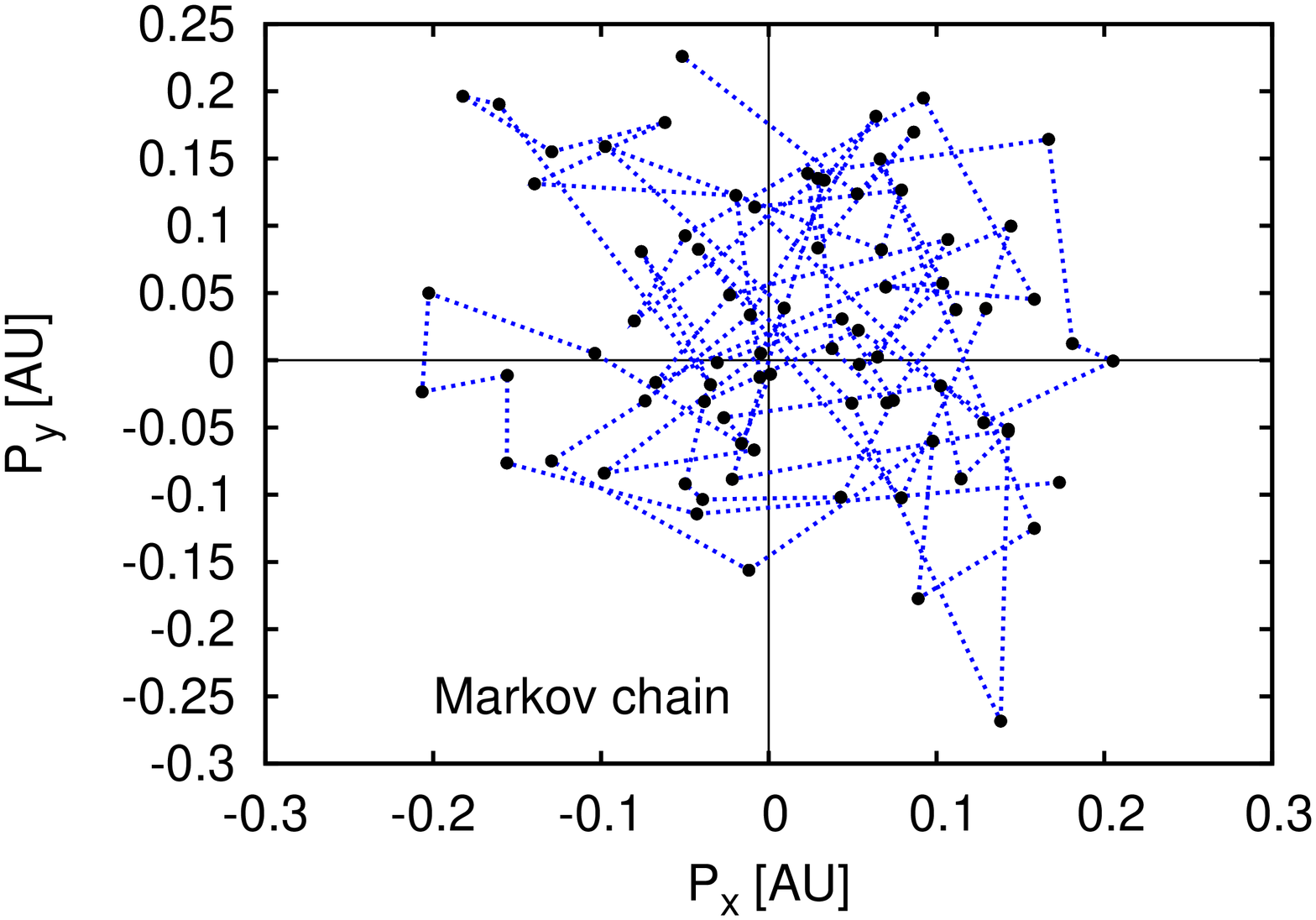}
   \caption{Upper panel: photocentre positions extracted from the snapshots over the five years of simulation. Lower panel: photocentre positions generated with a Markov chain with  $\sigma_P = 0.09$ AU and $\tau = 4 \times 10^6$s. Points are 23 days apart in both panels.}
   \label{generatedpositions}
\end{figure}
\newline
 We computed the correlation function for the sequence of photocentre positions extracted from the snapshots of the 3D model in the Gaia $G$ band shown in the top panel of Fig.~\ref{generatedpositions}. The result is shown as thick lines in Fig.~\ref{corrfunc}. 
Because of the time step of 23 days between two successive snapshots, we are only able to evaluate the correlation function on a discrete mesh with a 23-day time step. We see in Fig.~\ref{corrfunc} that the correlation function falls to zero for $\Delta t \approx 60$ days. Assuming the correlation function to be exponentially decreasing as in Eq.~(\ref{eq:corrfuncMc}), $\tau \simeq 46$ days $\simeq 4\times10^6$s. The variance $\sigma^2_P$ is taken as the arithmetic mean of $\sigma^2_{P,x}$ and $\sigma^2_{P,y}$, the values of the correlation function at $\Delta t=0$: $\sigma_P \simeq 0.09$ AU. To check whether the assumed exponential correlation function is compatible with the one extracted from the 3D model data, we generated photocentre positions with $\sigma_P= 0.09$ AU and $\tau=2-4\times10^6$s. An example of the generated positions is shown in the lower panel of Fig.~\ref{generatedpositions}. We then computed the correlation function with time-steps of 23 days. The results for some series of generated positions are shown as thin lines in Fig.~\ref{corrfunc} and are in good agreement with the correlation function for the 3D model points. Furthermore, it shows that we cannot distinguish between $\tau=2\times10^6$s and  $\tau=4\times10^6$s. \\
we note that the two parameters $\sigma_P$ and $\tau$ depend on the surface gravity and on the effective temperature of the star \citep{chiava2010}. This means that the values we derived here are only valid for stars with physical parameters similar to the ones of the 3D model. In the analysis below, we also consider stars with different $\sigma_P$ and $\tau$ in order to provide predictions for a wider class of objects. 
\newline
In Fig.~\ref{generatedpositionshisto}, we compare the distributions of $P_x$ and $P_y$ extracted from the snapshots of the 3D model and the ones generated with a Markov chain.
\begin{figure}[!h]
   \includegraphics[angle=270,width=\hsize]{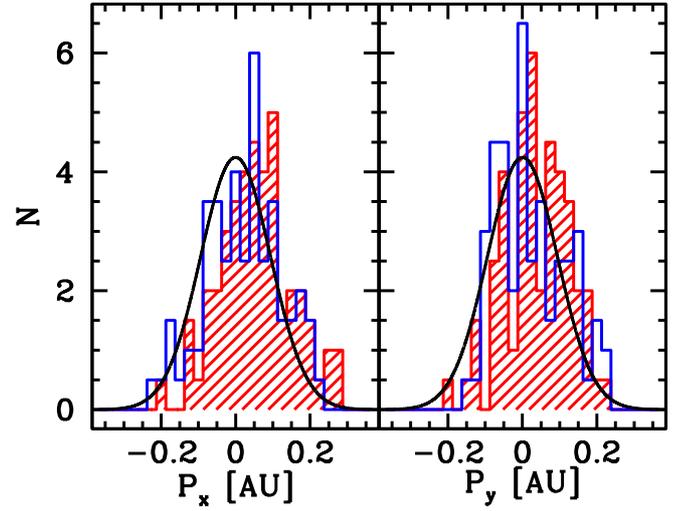}
    \caption{Histograms of the $P_x$ (left panel) and $P_y$ (right panel) positions extracted from the snapshots of the 3D model (red striped histograms) and the ones generated with the Markov chain (blue solid histograms). The actual $P_x$ and $P_y$ are shown in Fig.~\ref{generatedpositions}. The black solid line represents the Gaussian distribution used in the Markov chain to generate the photocentre positions.}
   \label{generatedpositionshisto}
\end{figure}
 The solid black line is the Gaussian distribution $N(0,\sigma_P)$ from which the generated positions are extracted. The 80 measurements considered here are not numerous enough to clearly reproduce the Gaussian distribution, though it does so if we simulate more photocentre positions with the Markov chain. We could not verify whether the distribution of the positions from the 3D model becomes Gaussian for longer times or with more measurements because no additional points are available.\\
The next step is to add these photocentre displacements generated with the Markov chain to simulated Gaia transit data.\\
The stellar population considered is a sample of $\simeq 5000$ red supergiants distributed around the Galactic plane ($0 \leq l \leq 180$ and $-20 \leq b \leq 20$ in Galactic coordinates). These stars have absolute $V$ magnitudes between -7 and -3.5 and apparent $G$ magnitudes between 5.6 and 20. Almost no stars are found at less than 1 kpc (for example see Fig.~\ref{parallax}). For such close distances, supergiants tend to appear brighter than the CCD saturation limit at $G=5.6$ and will not be observed. The sample was generated with the Besan\c{c}on model \citep{robi2004} and virtually observed with the Gaia Object Generator v7.0 (GOG; \citeauthor{isas2010} \citeyear{isas2010}). We then added to these ``observations'' photocentric shifts generated with a Markov chain. In this paper, we always add to all the stars in our sample a photocentric noise generated with Markov chains of the \textit {same} $\sigma_P$ and $\tau$ regardless of the physical parameters of the star.\\
The conclusions we draw in the analysis below shiuld be interpreted  in a statistical sense, as star-by-star effects are closely related to the particular Gaia scanning law and actual photocentre configuration.
\section{Impact on the single-star parameters}
As mentioned in the previous section, the goal of our work is to study the impact of surface brightness asymmetries on the stellar astrometric parameters derived by Gaia.
\begin{figure}[!h]
        \includegraphics[width=\hsize]{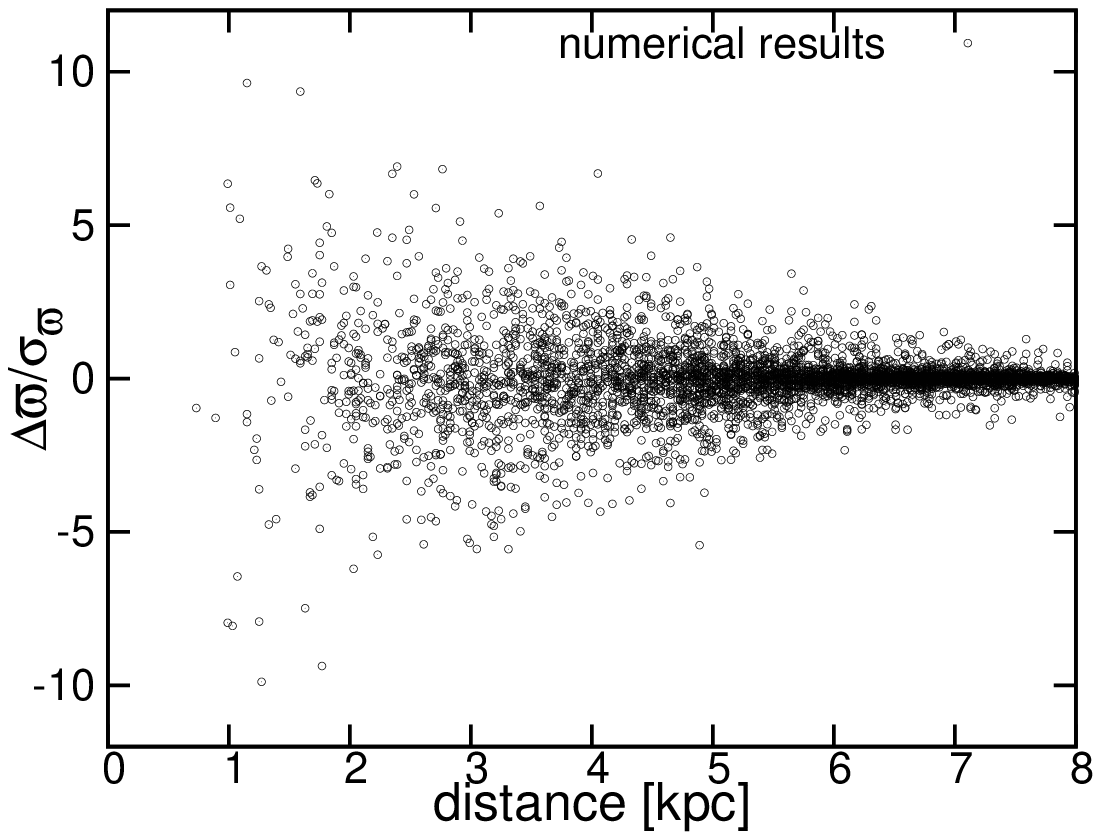}
\includegraphics[angle=270,width=\hsize]{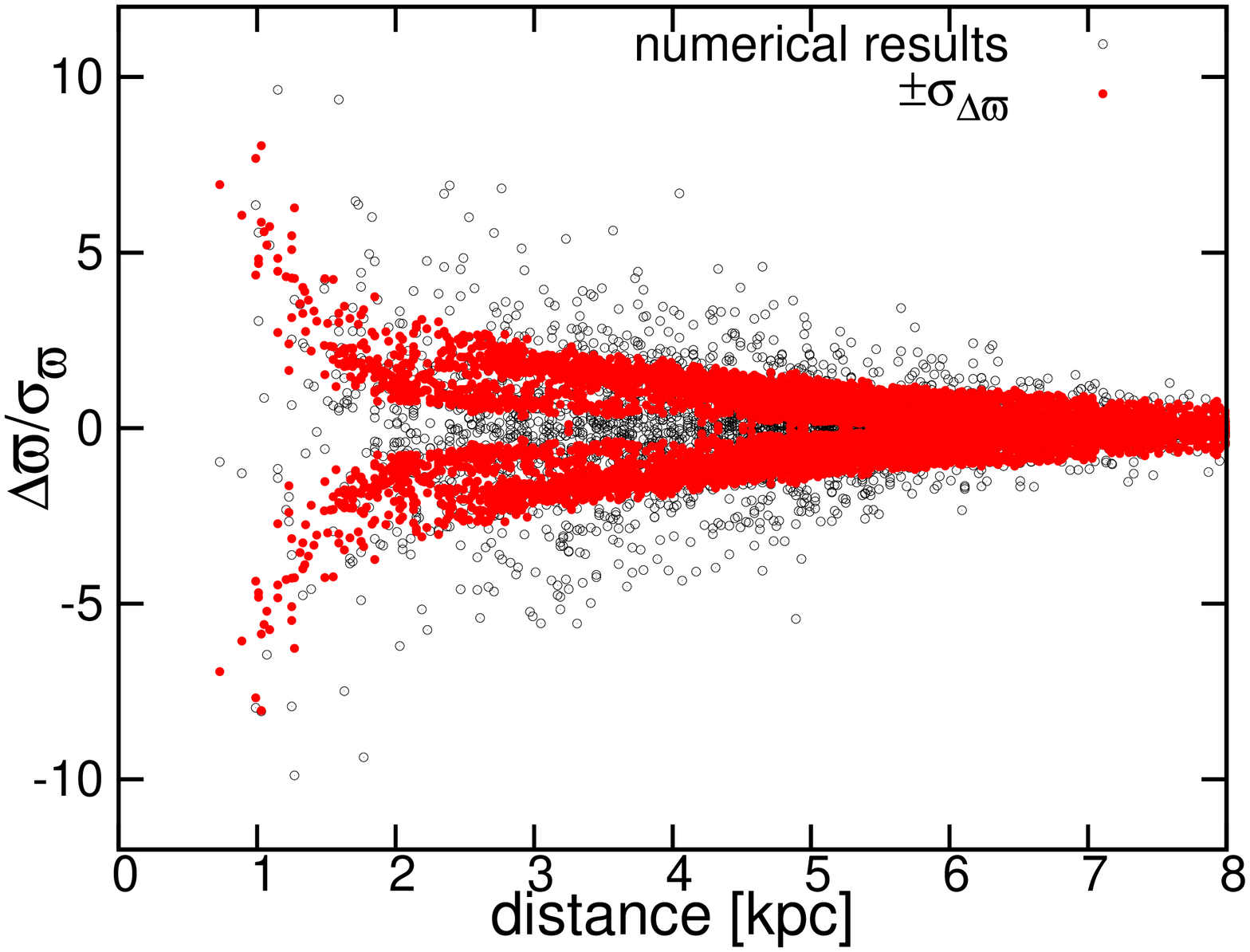}
      \caption{Upper panel: relative variation in the single-star parallax ($(\hat{\varpi}_{na}-\hat{\varpi}_{a})/\sigma_{\varpi}$) resulting from the addition of surface brightness asymmetries. Lower panel: same as upper panel but with the standard deviation predicted from Eq.~(\ref{eq:SD}) added (red points). The parameters of the Markov chain used to generate the data are $\sigma_P=0.09$ AU and $\tau = 4 \times 10^6$s.}
         \label{parallax}
   \end{figure}
The observations, as described in the previous section, are obtained  by adding photocentre motions generated with a Markov chain to Gaia transit data. They were fitted with the five-parameter single-star model ($\alpha_0^*$, $\delta_0$, $\varpi$, $\mu_{\alpha^*}$, $\mu_{\delta}$). In Fig.~\ref{parallax}, we show some results concerning the parallax. The upper panel of Fig.~\ref{parallax} illustrates the difference $\Delta \hat{\varpi}$ between the parallaxes determined with ($\hat{\varpi}_a$) and without ($\hat{\varpi}_{na}$) the surface brightness asymmetries, in units of the statistical error in the parallax  $\sigma_{\varpi}$. We note that $\sigma_{\varpi}$ will be provided as output in the Gaia catalogue and is derived from the covariance matrix of the parameters obtained in the data fitting process. As shown in Appendix~A of \cite{chiava2010} and of this paper, the quantity $\sigma_{\varpi}$ is independent of the presence of surface brightness asymmetries. The effect of the photocentre displacement can be quite significant as it can induce deviations in the derived parallaxes of up to 10$\sigma$ for $d\approx 1$ kpc. Nevertheless, we not that all effects become negligible for stars farther out than 5-6 kpc. At this distance, the epoch astrometric precision, which is about 0.03 mas for a star brighter than $G=12$, is comparable to the angular dimension of the added photocentre displacement (0.09 AU on average). A smaller photocentre noise will clearly have an impact only for more nearby stars, and the opposite for larger $\sigma_P$. For example, at $\sigma_P=0.02$ AU, only the most nearby stars ($d<2$ kpc) have a $\Delta \hat{\varpi}$ greater than 3$\sigma_{\varpi}$. On the other hand, at $\sigma_P=0.2$ AU, a deviation up to 30$\sigma$ is reached.\\
The expected value of $\Delta \hat{\varpi}$ can be easily computed (details in Appendix~A) assuming that the photocentre displacements are taken from a Gaussian distribution with variance $\sigma^2_P$. We note that with this approach the positions of the photocentre are treated as uncorrelated, which, as we have seen, is not strictly true. The mean value of $\Delta \hat{\varpi}$ is zero because the photocentre positions, on average, coincide with the geometric centre of the star. This is also true for correlated photocentre positions if enough of them are considered. On the other hand, $\sigma_{\Delta \hat{\varpi}}$, the standard deviation in $\Delta \hat{\varpi}$, indicates how much individual $\hat{\varpi}_a$ and $\hat{\varpi}_{na}$ values can differ:
\begin{equation}
\sigma_{\Delta\hat{\varpi}}=\displaystyle\sqrt{N\frac{\sigma^2_{P_{\alpha}}+\sigma^2_{P_{\delta}}}{2}}\frac{\sigma_P}{d}\frac{\sigma^2_{\varpi}}{\sigma_{w}^2},
\label{eq:SD}
\end{equation}
where $d$ is the distance in kpc, $N$ is the number of measurements, $\sigma_{w}$ is the epoch astrometric precision, $\sigma^2_{P_{\alpha}}$ and $\sigma^2_{P_{\delta}}$ are the variances in the parallax factors \citep[and Eqs.~\ref{eq:sigmaPa} and \ref{eq:sigmaPd}]{1985green}. 
In the lower panel of Fig.~\ref{parallax}, we have plotted the $\sigma_{\Delta \hat{\varpi}}$ values estimated from Eq. (\ref{eq:SD}) on top of the $\Delta \hat{\varpi}$ resulting from the single-star astrometric fit. The agreement between the two sets of points is remarkably good, meaning that $\sigma_{\Delta \hat{\varpi}}$ can be used as a realistic estimate of the impact of the surface brightness asymmetries on the accuracy of $\hat{\varpi}_a$.\\
The Gaia end-of-mission astrometric accuracy $\sigma_{\varpi}$ for a ``perfect'' single star can be written as \citep{debruijne2005}
\begin{equation}
\sigma_{\varpi}=m\, g_{\varpi} \frac{\sigma_w}{\sqrt{N}}.
\label{eq:endofmissionerror}
\end{equation}
In this expression, $m$ is an overall, end-of-mission contingency margin ($m=1.2$ for a 20\% margin), $g_{\varpi}$ is a dimensionless geometrical factor relating to the scanning geometry, $\sigma_w$ is the single measurement error (including all the error sources), and $N$ is the number of transits. We can rewrite Eq.~(\ref{eq:SD}) for $\sigma_{\Delta \hat{\varpi}}$ in terms of Eq.~(\ref{eq:endofmissionerror})
\begin{equation}
\sigma_{\Delta \hat{\varpi}}= \frac{m^2 g_{\varpi}^2}{\sqrt{N}}\frac{\sigma_P}{d}\sqrt{\frac{\sigma^2_{P_{\alpha}}+\sigma^2_{P_{\delta}}}{2}}\propto\frac{\sigma_{\Theta}}{\sqrt{N}}.
\label{eq:gendepdeltabeta}
\end{equation}
We see that $\sigma_{\Delta \hat{\varpi}}$ is only related to the angular size of the photocentre displacement induced by the surface brightness asymmetries ($ \sigma_{\Theta}=\frac{\sigma_P}{d}$) and the scanning geometry. There is no dependence on the instrumental sensitivity $\sigma_w$ because $\sigma_{\Delta \hat{\varpi}}$ is intrinsic to the physics of the star. On the other hand, $\frac{\sigma_{\Delta \hat{\varpi}}}{\sigma_{\varpi}}$, the capability of Gaia to detect the noise induced by the surface brightness asymmetries, does depend on $\sigma_w$ such that
\begin{equation}
\frac{\sigma_{\Delta \hat{\varpi}}}{\sigma_{\varpi}}\propto \frac{\sigma_{\Theta}}{\sigma_w}.
\label{equno}
\end{equation}
For red supergiants, $\sigma_w$ is roughly constant because $G< 12.6$ in most of the cases.\\
Finally, if we consider $\frac{\sigma_{\Delta \hat{\varpi}}}{\varpi}$, which is the parallax inaccuracy (owing to the surface brightness asymmetries) relative to the parallax, we find that it is independent of the distance of the star
\begin{equation}
\frac{\sigma_{\Delta \hat{\varpi}}}{\varpi}\propto\sigma_P.
\label{eqdue}
\end{equation}
We can now compute $\frac{\sigma_{\Delta \hat{\varpi}}}{\sigma_{\varpi}}$ and  $\frac{\sigma_{\Delta \hat{\varpi}}}{\varpi}$ for a red supergiant star using mean values for the scanning-law parameters $g_{\varpi}=1.91$, $m=1.2$, $N=78$, $\sigma_w=30$ $\mu$as \citep{debruijne2005, 2009lindegren}, and $\sigma_P=0.09$ AU. For supergiants located at 1 kpc ($\sigma_{\Theta}=0.09$ mas), Eq.~(\ref{equno}) predicts that $\frac{\sigma_{\Delta \hat{\varpi}}}{\sigma_{\varpi}}<4.3$ for 68\% of the sample and values larger than 13 for 0.3\% of the sample. For supergiants at 5 kpc ($\sigma_{\Theta}=0.018$ mas), these numbers are 0.8 and 2.4, respectively. 
Similarly, Eq.~(\ref{eqdue}) predicts that $\frac{\sigma_{\Delta \hat{\varpi}}}{\varpi}<3.4\%$ for 68\% of the supergiants (independently of their distance) and values larger than 10.3\% for 0.3\% of the sample.\\
A similar behaviour is found for the four other parameters ($\alpha_0^*$, $\delta_0$, $\mu_{\alpha^*}$, and $\mu_{\delta}$). Expressions equivalent to Eq.~(\ref{eq:SD}) are derived in Appendix~A. we note that for position and proper motion, we find values that are up to 15$\sigma$ off, while the photocentric noise was generated with the same Markov chain parameters as in Fig.~\ref{parallax}.\\
We now analyse the accuracy $\varpi_{true}-\hat{\varpi}_a$ (as opposed to $\Delta \hat{\varpi}=\hat{\varpi}_a-\hat{\varpi}_{na}$) of the ``fitted single-star'' parallax with respect to the true parallax.
\begin{figure}[!h]
        \includegraphics[angle=270, width=\hsize]{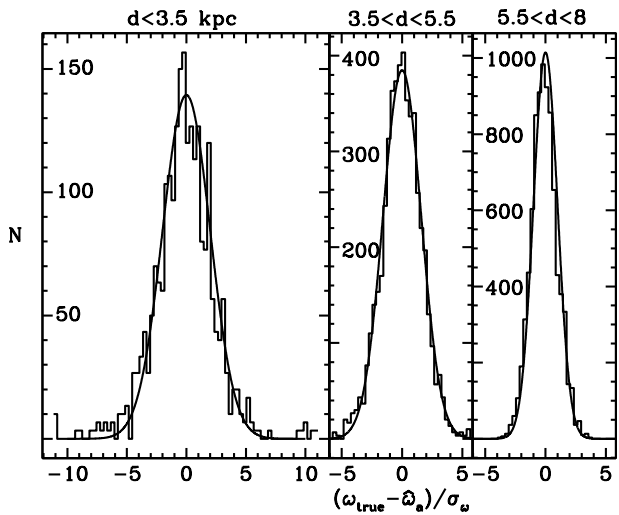}
      \caption{Histogram of the relative error in the single-star parallax (($\varpi_{true}-\hat{\varpi}_{a})/\sigma_{\varpi}$) resulting from the addition of surface brightness asymmetries for supergiant stars at different distances. The distribution for stars in the right panel ($d>5.5$ kpc) closely reproduces the expected $N(0, 1)$. In the central panel ($3.5<d<5.5$ kpc), the distribution is wider with a best-fit $N(0, 2.5)$. Finally, for the left panel ($d<3.5$ kpc) the $\sigma$ of the distribution increases to four. Parameters of the Markov chain used here are $\sigma_P=0.09$ AU and $\tau = 4 \times 10^6$s. The x-axes in the three panels are at the same scale.}
         \label{histotrueparallax}
   \end{figure}
In the three panels of Fig.~\ref{histotrueparallax}, we plot this difference, normalised by the estimated error in the parallax $\sigma_{\varpi}$, for different slices in distance. For stars far away (right panel), the distribution is clearly $N (0,1)$ as expected. As we consider more nearby stars, the distribution becomes wider and wider. For stars in the range $3.5<d<5.5$ kpc (central panel), the histogram is approximated well by a $N(0, 2.5)$ distribution, and for stars with $d<3.5$ kpc (left panel) by $N(0, 4)$. If we increase $\sigma_P$, we obtain wider and wider kpc distributions. For $\sigma_P=0.2$ AU, only stars between 7.5 kpc and 8 kpc follow a $N(0, 1)$ distribution. On the other hand, lowering $\sigma_P$ produces narrower distributions. For $\sigma_P=0.02$ AU, the distribution becomes $N(0, 1)$ for stars at all distances indicating that at this level of photocentric noise, no effects are  expected in Gaia data processing. We recall that actual stars have $\sigma_P$, which is dependent upon their absolute magnitude, but the exact relationship is not known so far \citep[]{chiava2010}.

\begin{figure}[!h]
    \centering \includegraphics[angle=270,width=\hsize]{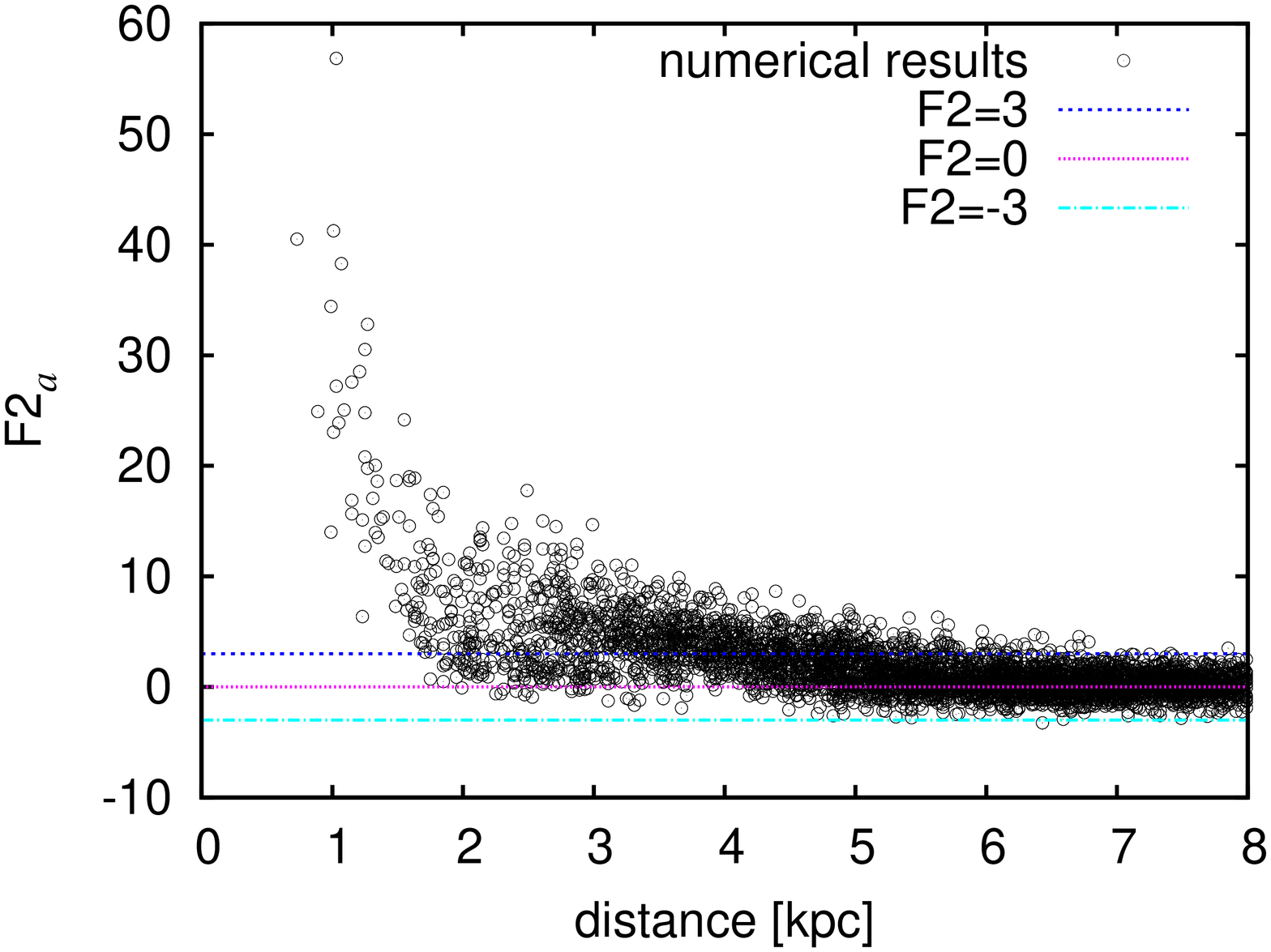}
 \centering \includegraphics[angle=270,width=\hsize]{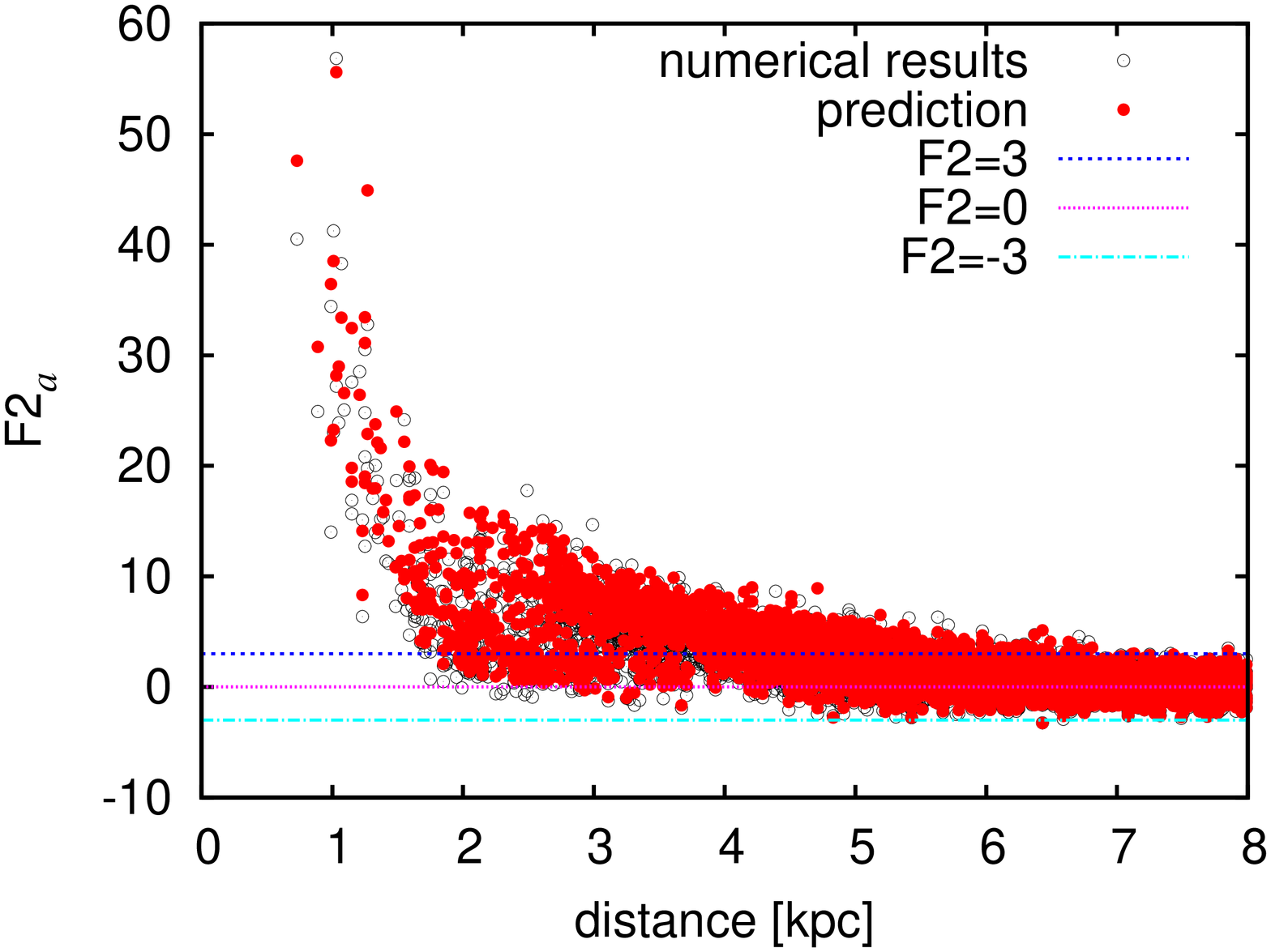}
      \caption{Upper panel: $F2$ for the single-star fit of the simulated data with the addition of surface brightness asymmetries. A $F2>3$ indicates a non-satisfactory fit. Lower panel: the same as the upper panel but with the theoretical prediction (Eq.~(\ref{eq:F2})) added on top (see text).}
         \label{GoF}
   \end{figure}

As already mentioned, we have used a five-parameter-model for our set of supergiants because we know we are dealing with genuine single stars. However, the $F2$ values for these solutions (Fig.~\ref{GoF}, top panel) can exceed the threshold value three by more than one order of magnitude. An estimate of the increase in $F2$ induced by the presence of surface brightness asymmetries can be easily obtained by noticing that $\chi^2_{a}$, which is the $\chi^2$ in the presence of surface brightness asymmetries, can be written as \citep[Eqs.~11 and 12, and Appendix B of this paper]{chiava2010}
\begin{equation}
\chi^2_{a}\simeq \chi^2_{na}+ N\left(\frac{\sigma_P}{d\cdot \sigma_{w}}\right)^2,
\label{eq:chiquadro}
\end{equation}
where $\chi^2_{na}$ is the  $\chi^2$ of the fit of the data without surface brightness asymmetries, $N$ is the number of measurements, $d$ is the distance in kpc, and $\sigma_{w}$ is the Gaia single-measurement error. This expression can be easily converted into the analogous expression for $F2$
\begin{equation}
F2_{a}=F2_{na}+\left(\frac{9\nu}{2}\right)^{1/2}\left[\left(1+\frac{\sigma_P^2}{\sigma_w^2}\right)^{1/3}-1\right].
\label{eq:F2}
\end{equation}
This quantity is plotted in the lower panel of Fig.~\ref{GoF} on top of the results obtained for the single-star solution.
There is on average remarkable agreement. As we previously pointed out, a case by case agreement is not to be expected. When we change the parameters of the Markov chain and set $\sigma_P=0.2$ AU, $F2_a$ increases to 100 for more nearby stars. No stars at $d< 2.5$ kpc have $F2_a<3$, and we find values larger than three at all distances. In contrast if we look at what happens with $\sigma_P=0.02$ AU, $F2_a$ becomes larger than three only for a negligible fraction of the stars, and it never exceeds seven.

\section{Non-single-star processing}
\begin{figure}[!h]
    \centering 
\includegraphics[angle=270,width=\hsize]{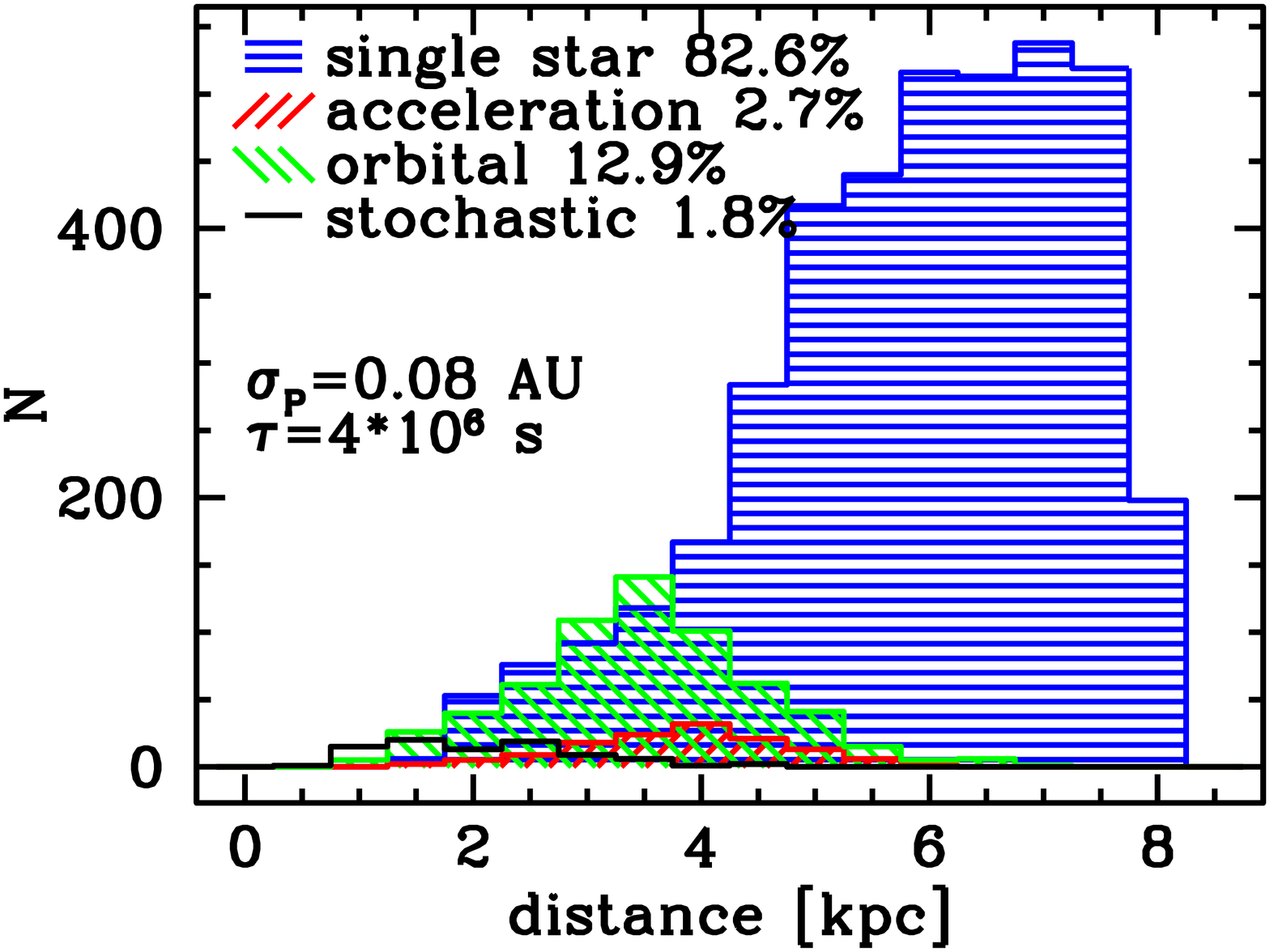}
   \includegraphics[angle=270,width=\hsize]{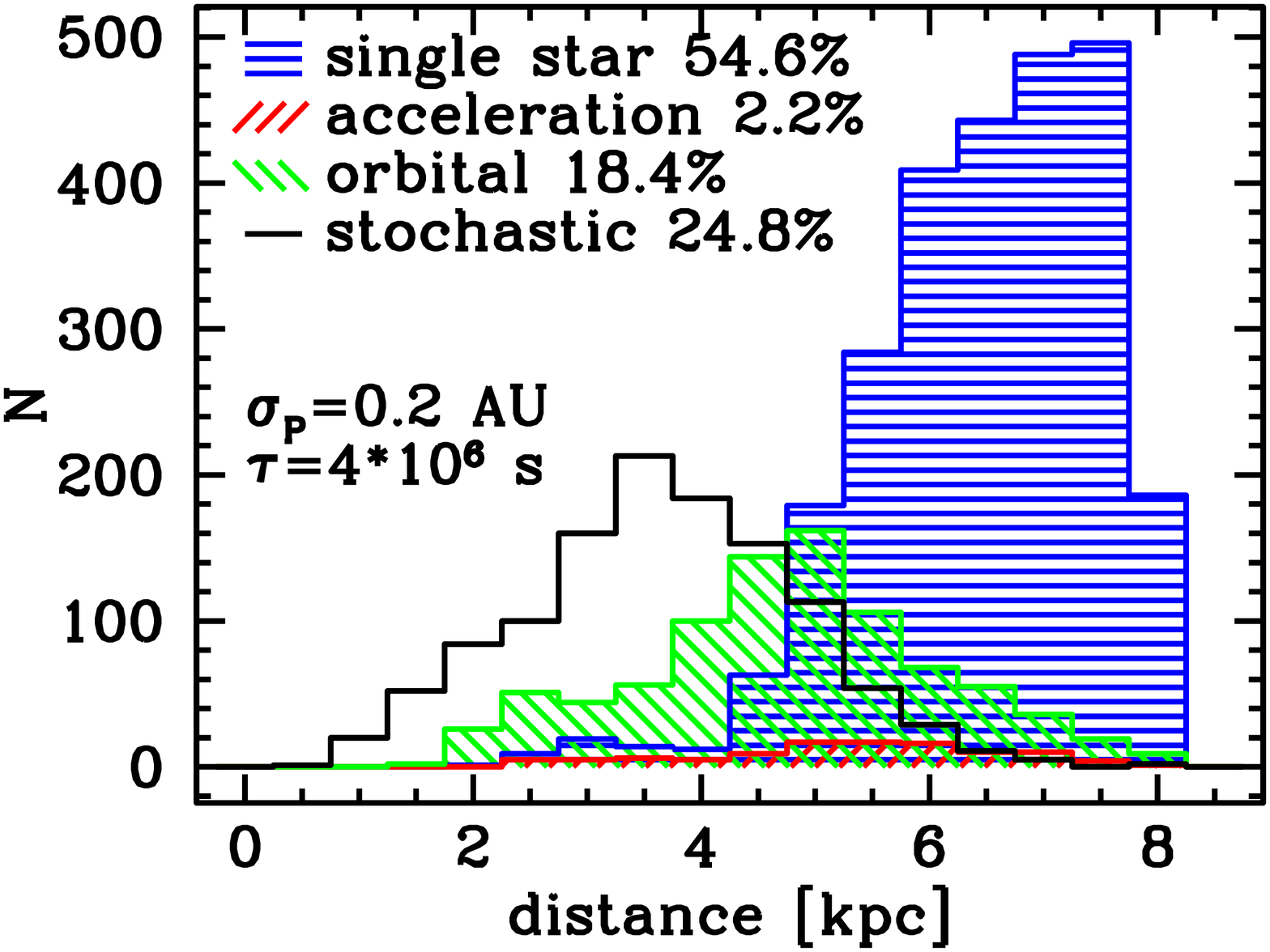}
   \includegraphics[angle=270,width=\hsize]{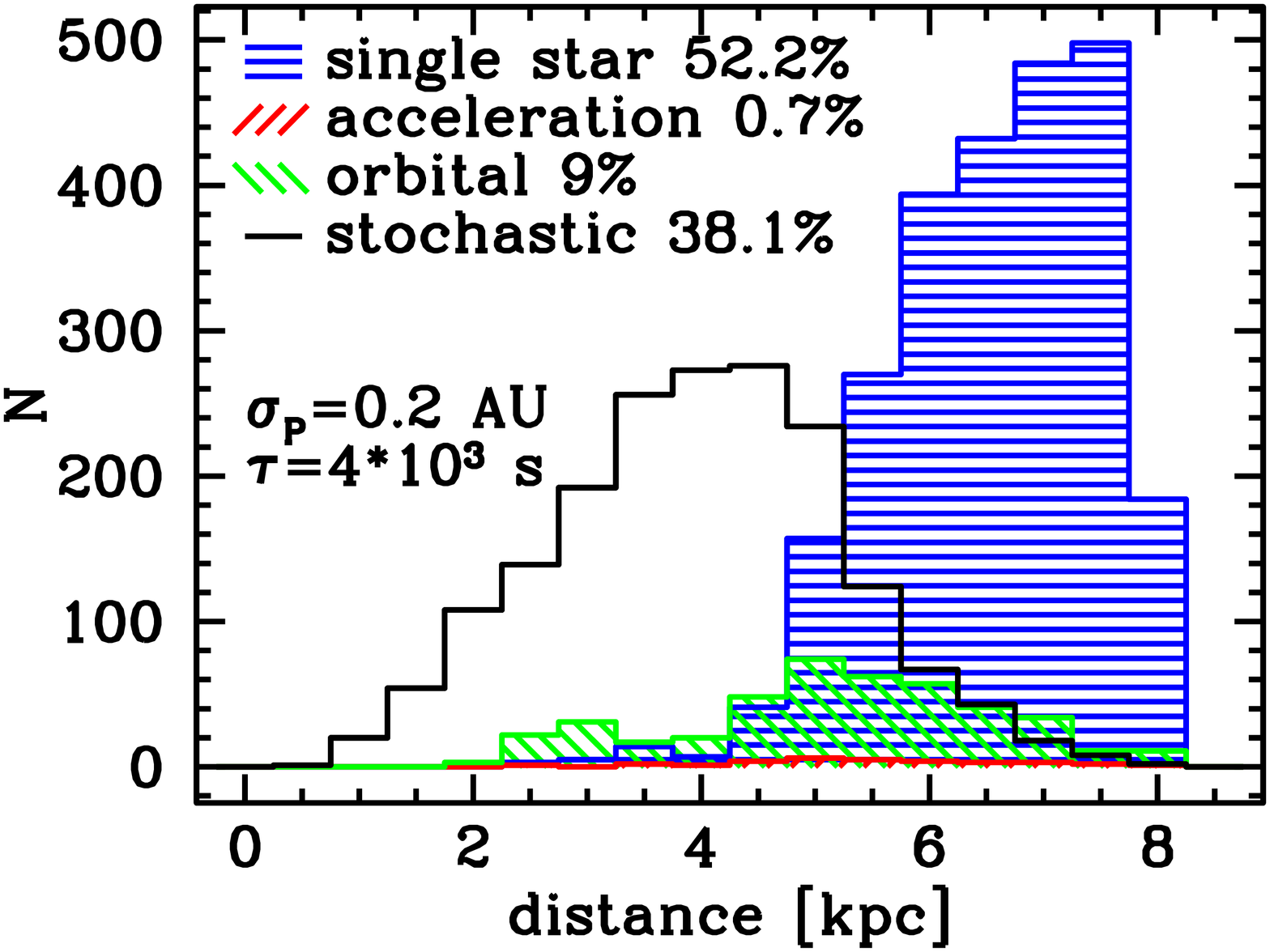}
      \caption{Distribution of the different types of solutions found as a function of the distance. In the three panels are shown the results for the same sample of stars to which different photocentric motions have been applied, as indicated by the corresponding $\sigma_P$ and $\tau$ in the figure label.}
         \label{istocol}
   \end{figure}

\begin{figure}[!h]
    \centering \includegraphics[width=0.495\hsize]{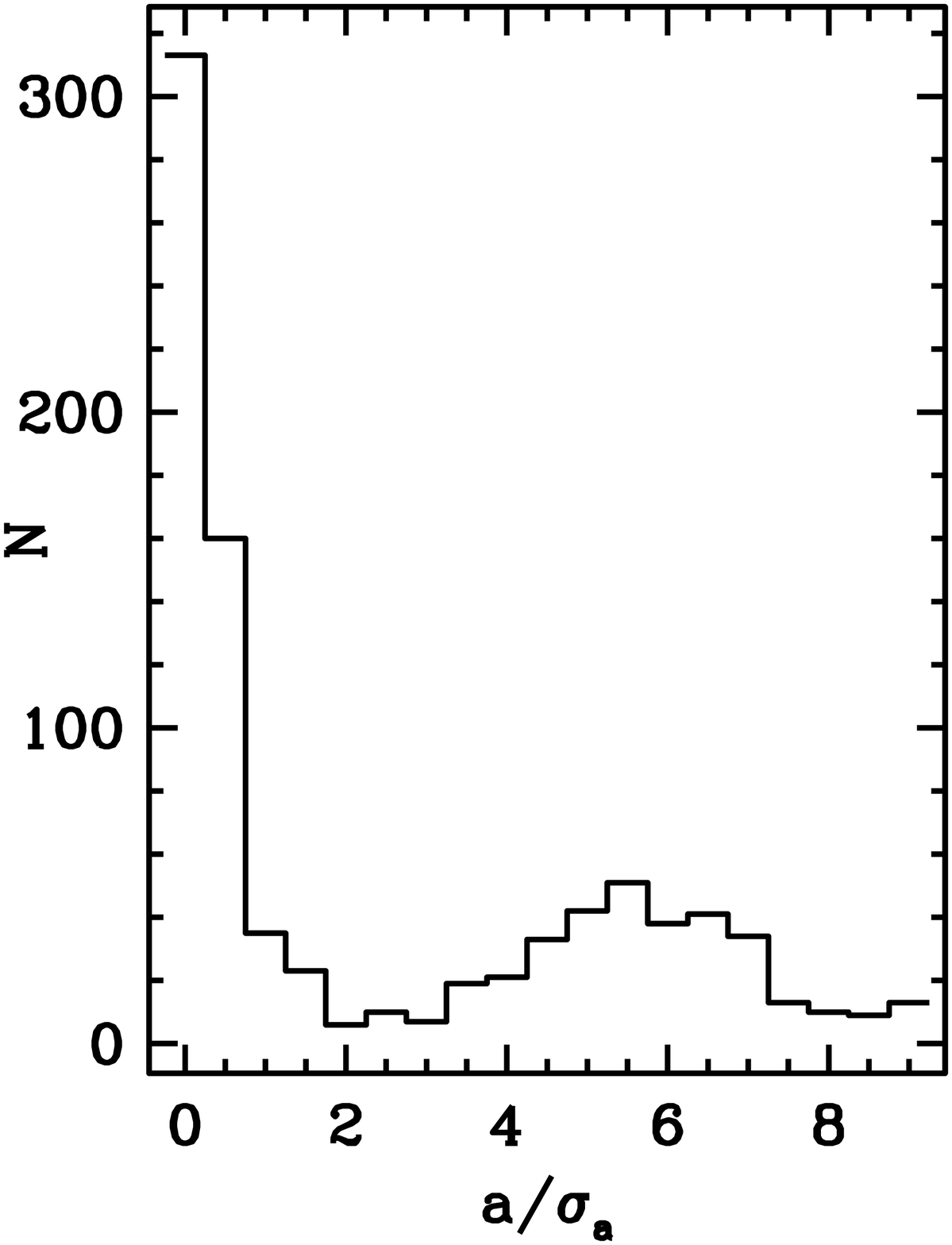}
  \centering \includegraphics[width=0.495\hsize]{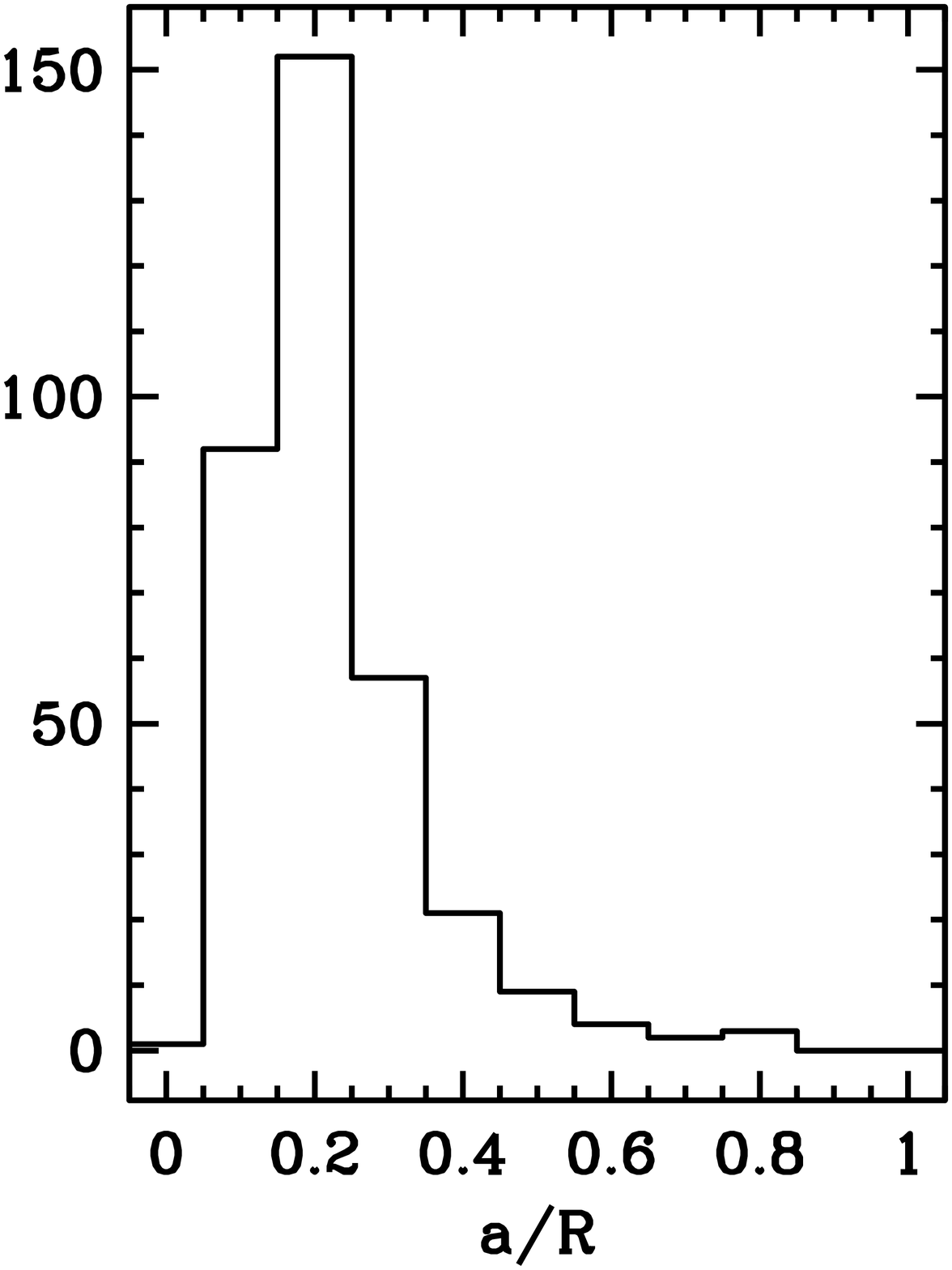}
      \caption{Left panel: distribution of the ratio $a/\sigma_a$, with $a$ the semi-major axis of the orbit and $\sigma_a$ its uncertainty. Solutions with $a/\sigma_a<2$ are not significant and may be rejected. Right panel: ratio $a/R$, with $R$ the stellar radius for stars with $a/\sigma_a>2$. All orbits appear to have a semi-major axis smaller than the stellar radius and can thus be rejected. Markov chain parameters:$\sigma_P = 0.2$ AU and $\tau = 4 \times 10^6$s }
         \label{istoa}
   \end{figure}

\begin{figure}[!h]
    \centering \includegraphics[angle=270,width=\hsize]{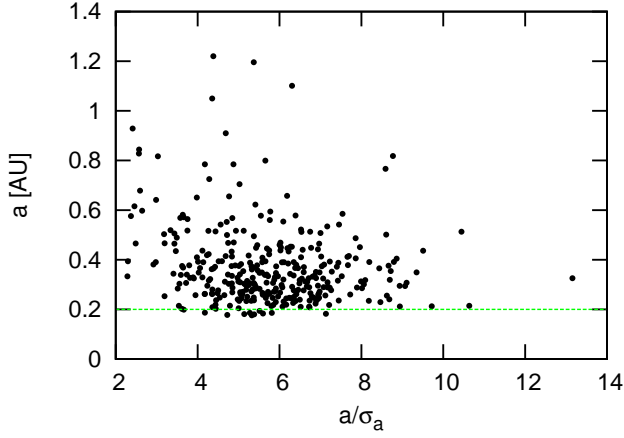}
      \caption{Semi-major axes $a$ of the orbits derived with the orbital solution. The physical size of $a$ in AU is comparable to the photocentre motion added to the simulated Gaia data. Markov chain parameters:$\sigma_P = 0.2$ AU and $\tau = 4 \times 10^6$s }
         \label{dista}
   \end{figure}

We have so far treated the supergiants in our sample as single stars, as we know that they actually are. Nevertheless, when processing real Gaia data, we will not know in advance the kind of object we will be analysing. 
When we do not restrict the data processing flow to the five-parameter solution, stars with a single-star $F2$ larger than three are further processed. A chain of solutions of increasing complexity is applied until $F2$ becomes smaller than three. In the acceleration solution, a possible variation of the proper motion is added to better fit the data. In the orbital solution, the parameters of the orbit of a binary system are added. If neither the single-star solution nor the non-single-star solutions (acceleration or orbital) are acceptable on the basis of their $F2$, a stochastic solution is computed: some noise is added to the along-scan measurements in order to lower to zero the $F2$ of the single-star solution.\\ 
For our sample, an acceleration solution, an orbital solution, or a stochastic solution are found for some fraction of the stars, and this fraction depends on the parameters of the Markov chain. The distribution of solutions among these possible classes is plotted in Fig.~\ref{istocol} for three different sets of $\sigma_P$ and $\tau$.
The number of single-star solutions decreases drastically as we increase $\sigma_P$. Moreover, the impact on the astrometric parameters then increases, implying that the number of solutions in which they are unaltered decreases. We see as well that all the nearby ($d\lesssim 4$ kpc) single-star solutions disappear. At the same time, we see an order-of-magnitude increase in the number of stochastic solutions. Finally, we point out the decrease of the fraction of orbital solutions as we decrease $\tau$. This is a consequence of the loss of correlation between temporally close measurements that makes it harder to fit the points with an orbit.\\
That single-star solutions are rejected by the Gaia pipeline even though we are dealing with genuine single stars is unfortunate. Hence, we analysed these solutions to see whether it is possible to identify them and filter them out. We focused our attention on the orbital solutions as they can represent a significant fraction of the solutions for our sample.\\
As a first step, we analysed the significance of the orbital solution by looking at the ratio $a/\sigma_a$, where $a$ is the semi-major axis of the orbit. If $a$ is smaller or comparable to its error $\sigma_a$, the solution is not significant as this is an indication of null orbital parameters. In the left panel of Fig.~\ref{istoa}, we show the ratio  $a/\sigma_a$ for a Markov chain with $\sigma_P = 0.2$ AU and $\tau = 4 \times 10^6$s and conclude that a large fraction of these solutions has $a/\sigma_a<1$. They are thus spurious and may easily be flagged as such. For these solutions, the reduction in $F2$ under the acceptance value of 3 is caused by the $\chi^2$ decreasing faster than the number of parameters increases.
As a second step, we studied the stars for which the orbital solution is significant (i.e. located in the peak at about $a/\sigma_a=6$ in the left panel of Fig.~\ref{istoa}) and we computed the ratio $a/R$, where $R$ is the radius of the star. This quantity is plotted in the right panel of Fig.~\ref{istoa} and is always smaller than unity. We may thus reject all those solutions as they are unphysical, representing orbits with separations smaller than the radius of the stars under study. 
Moreover, we note that the size in AU of the semi-major axis found (Fig.~\ref{dista}) is always of the same order of magnitude as the photocentre motion added with the Markov chain, 0.2 AU being about the minimum value found.

\begin{figure}[!h]
     \centering \includegraphics[angle=270,width=\hsize]{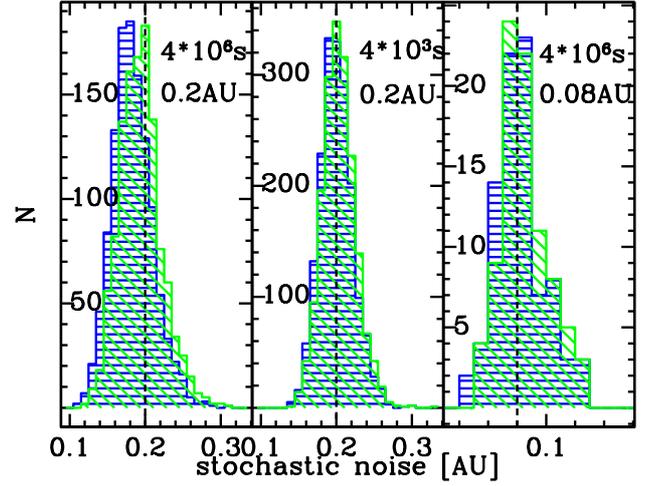} 
      \caption{The blue horizontally striped histograms show the ``stochastic'' noise added to Gaia measurements in the computation of a stochastic solution. The green diagonally striped ones represent the stochastic noise plus the offset of the mean photocentre position with respect to the centre of the star. The three panels show results for three different sets of parameters of the Markov chain. The vertical dashed line corresponds to the standard deviation $\sigma_P$ in the photocentre displacement.} 
         \label{istostoc}
   \end{figure}

Finally, we discuss the stochastic solutions. Stochastic noise is meant to represent any source of ``noise'' found in the data that cannot be modelled with any of the other models. We expect it to be correlated with the noise we added to the single-star Gaia data. A histogram of the stochastic noise (in AU) is shown in Fig.~\ref{istostoc} for three different sets of parameters of the Markov chain. We show examples with extreme parameters in order to more clearly illustrate the occurring effects. In the left panel, we show the results for simulated photocentric motions with $\sigma_P=0.2$ AU and $\tau=4\times 10^6$s. The  blue horizontally striped histogram represents the distribution of the stochastic noise obtained by the data processing. It is clearly neither symmetric nor centred at the value of $\sigma_P=0.2$ AU as expected (vertical dashed line). This is because in the fitting process, the position of the star is assumed to be at the mean photocentre position, which does not necessarily coincide with the geometrical centre of the star (example in the top panel of Fig.~\ref{generatedpositions}, where the mean photocentre position is in the first quadrant rather than at the origin). Once we quadratically add to the stochastic noise the difference between the mean photocentre position and the true position of the star, the resulting distribution (green diagonally striped histogram) becomes centred at 0.2 AU as expected. This positional shift and the asymmetry of the distribution is caused by the length of the Gaia mission (5 years) not being long enough for the distribution of photocentre positions to be Gaussian. Moreover,  that measurements are taken at time intervals shorter than the relaxation time of about 40 days and the non-uniformity of the time sampling play a crucial role.
The minimum time between two consecutive Gaia observations of the same target is about two hours. When we decrease $\tau$ to $4\times 10^3$s (i.e. about one hour, meaning that all measured photocentre positions are almost uncorrelated), the histogram becomes almost symmetric and centred on 0.2 AU (central panel of Fig.~\ref{istostoc}). If we keep $\tau=4\times 10^6$s and decrease  $\sigma_P$ to 0.08 AU (right panel of Fig.~\ref{istostoc}), the stochastic noise distribution is still asymmetric and not peaked at 0.08 AU. However, the effects are mainly smaller than for the 0.2 AU case, merely because of the smaller size of the photocentric shifts generated. 
  
From this discussion, we can conclude that the stochastic noise can be used as an estimate of the photocentric noise present on the surface of the star. This correspondence is at its closest for short correlation times and small photocentric motions. Under these circumstances, the distribution will be more symmetric and will peak  at the correct value.  

\section{Conclusions} 
Surface brightness asymmetries are a common characteristic of stars and cause time-varying positional shifts between the photocentre and the barycentre. We have analysed their effects on the parameters derived from Gaia simulated data for supergiants. To achieve this goal, on the basis of a 3D model of a supergiant, we have simulated photocentric displacements to be added to along-scan measurements. We have then performed a five-parameter fit.
For such a single-star solution, we have shown that the deduced parallaxes
can be up to $13 \sigma_{\varpi}$ off for a supergiant at 1~kpc with $\sigma_P = 0.09$~AU and
$\frac{\sigma_{\Delta\hat{\varpi}}}{\varpi}$  up to 10\%. We have shown as well that
 $\frac{\sigma _{\Delta\hat{\varpi}}}{\sigma_{\varpi}}$ decreases with distance as $\frac{\sigma_P}{d \sigma_w}$,
but in contrast,  $\frac{\sigma_{\Delta\hat{\varpi}}}{\varpi}$ does not vary with distance, being
 proportional to $\sigma_P$. At the same time, the quality of the fit can be very poor, with $F2$ parameters exceeding the threshold value of 3 by more than one order of magnitude. If more complex solutions are searched for, even though $F2$ becomes smaller than three, there is no improvement in the accuracy of the derived parameters and, at the same time, the computational effort is substantially increased. Spurious binary solutions can be rejected on the basis of significance tests and the comparison of the size of the semi-major axis with the radius of the star. When a stochastic solution is found, we observed a correlation between the stochastic noise and the Gaussian noise injected in the Markov chain to simulate the photocentric motion. This information could be exploited to obtain, at least, the order of magnitude of the convective noise present.\\
From the standpoint of the Gaia data processing, if a star could be identified
as a supergiant independently of its parallax, a 'supergiant flag' may then be used
as a warning that the astrometric parameters may be far off.
Classification methods independent of the parallax include the spectroscopic determination
of the atmospheric parameters ($T_{\rm eff}, \log g$) from the Gaia spectrometer, or the
classification as an irregular variable from the repeated photometric measurements.
Irregular variables exist in two flavours, however: supergiants and stars at the
base of the red giant branch \citep{Jorissen-1997}.
The two classes of irregular variables differ by several magnitudes in their respective
absolute magnitudes, so that a 10\% relative error in the parallax of the
supergiant caused
by the surface brightness asymmetries is not large enough to intermingle the two classes and
 introduce confusion.\\
To identify true binaries among supergiants despite the confusion
introduced by the surface brightness asymmetries, an indirect way
is provided by the detection of composite colour indices (in addition to 
the orbital separation probably now being significantly larger than the stellar radius). Such
composite colour indices for systems involving supergiants were  computed by
\citet{2011malkov}, with the conclusion that these composite indices differ by at least
0.3~mag from the nearest single-star indices in the best two-colour diagram, thus allowing the identification of the
binary systems.\\
This complementary non-astrometric information should thus allow for a more efficient and
reliable evaluation of the significance of the astrometric solution in the presence of
surface brightness asymmetries.\\

\begin{acknowledgements}
E.P. is supported by the ELSA (European Leadership in Space Astrometry) Research
Training Network of the FP6 Programme and PRODEX. D.P. is FNRS Senior Research Associate. We thank L.Lindegren, B.Holl, A. Chiavassa, and C. Babusiaux for their support and useful discussions. Simulated data provided by the Simulation Unit (CU2) of the Gaia Data Processing Analysis Consortium (DPAC) have been used to complete this work. They are gratefully acknowledged for this contribution.
\end{acknowledgements}

\bibliographystyle{aa}
\bibliography{biblio}

\begin{appendix}
\section{Estimation of the accuracy on the astrometric parameters}
Gaia will provide $N$ measurements ($\vec{w}$) of the along-scan position of each star and hence the astrometric solution will be computed by solving an overdetermined system of $N$ equations for $\vec{\beta_{na}}$, the $n$ unknowns ($\alpha_0^*$, $\delta_0$, $\varpi$, $\mu_{\alpha^*}$, $\mu_{\delta}$ in the case of a single-star solution discussed here):
\begin{equation}
X\vec{\beta_{na}}+\vec{\epsilon}=\vec{w},
\label{eq:hdix}
\end{equation}
where $\vec{\epsilon}$ is the vector of errors that we assume to be random and mutually uncorrelated, i.e. they have zero mean and their covariance matrix $W$ is diagonal. We define $X$ to be the matrix of known coefficients
\begin{equation}
X=
\left(
\begin{array}{c c c c c}
\sin \theta_1 & \cos \theta_1 & f_{w_1} & t_1\sin \theta_1 & t_1\cos \theta_1 \\
\sin \theta_2 & \cos \theta_2 & f_{w_2} & t_2\sin \theta_2 & t_2\cos \theta_2 \\
\vdots        & \vdots        & \vdots  & \vdots         &\vdots \\
\sin \theta_N & \cos \theta_N & f_{w_N} & t_N\sin \theta_N & t_N\cos \theta_N  
\end{array}
\right),
\label{eq:Xdef}
\end{equation}
where $\theta$ is the position angle of the scan, $t$ is the observing time, and $f_w=r(P_{\alpha}\sin \theta+P_{\delta}\cos \theta)$ is a linear combination of the parallax factors $ P_{\alpha}$ and $P_{\delta}$  \citep{1985green}, and $r$ is the radius vector of Gaia's orbit at time $t$ in $AU$
\begin{equation}
P_{\alpha}=\cos \epsilon \cos \alpha \sin \odot - \sin \alpha \cos \odot,
\end{equation}
$
P_{\delta}= (\sin \epsilon \cos \delta - \cos \epsilon \sin \alpha \sin \delta)\sin \odot - \cos\alpha \sin \delta \cos \odot
$,
\bigskip
\\
where $\odot$ and $\epsilon$ are, respectively, the longitude of the Sun and the obliquity of the ecliptic, both at time $t$.\\
We wvaluate $\vec{\hat{\beta}_{na}}$, the best estimate of the parameters $\vec{\beta_{na}}$ in Eq.(\ref{eq:hdix}), by minimizing $\chi^2$
\begin{equation}
\chi_{na}^2=(\vec{w}-X\vec{\beta_{na}})^TW^{-1}(\vec{w}-X\vec{\beta_{na}}),
\label{eq:chi}
\end{equation}
which can be expressed by 
\begin{equation}
\vec{\hat{\beta}_{na}}=(X^TW^{-1}X)^{-1}(X^TW^{-1}\vec{w}).
\label{eq:betaest}
\end{equation}
The difference between the true parameters and the estimated ones is easily obtained using Eqs.~(\ref{eq:betaest}) and (\ref{eq:hdix})
\begin{equation}
\vec{\hat{\beta}_{na}}-\vec{\beta_{na}}=(X^TW^{-1}X)^{-1}X^TW^{-1}\vec{\epsilon},
\end{equation}
as well as $Z$, the covariance matrix of the parameters
\begin{equation}
Z=E[(\vec{\hat{\beta}_{na}}-(\vec{\beta_{na}}))(\vec{\hat{\beta}_{na}}-(\vec{\beta_{na}}))^T]=(X^TW^{-1}X)^{-1}.
\label{eq:covmat}
\end{equation}
We note that $Z$ is diagonal because all the off-diagonal terms contain $\sin \theta$, $\cos\theta$, or $t$ factors that averages to zero. The mean parallax factors are also zero because of their dependence on $\odot$.\\
%
%
We now consider a star with time-varying surface brightness asymmetries. Their presence causes a shift in the photocentre position and a consequent shift ($\Delta w$) in the observed along-scan position with respect to a star with a uniform disk.\\
This can be expressed as
\begin{equation}
\Delta w_i=\frac{1}{d} \left[ a_i \cos \theta_i + b_i \sin \theta_i \right]= \frac{1}{d} D_i,
\label{eq:defdeltaw}
\end{equation}
where $(a_i, b_i)$ are the coordinates of the photocentre in a reference system with its origin at the centre of the star (expressed in AU) and $d$ is the distance of the star in kpc. We assume that $a_i$ and $b_i$ follow a Gaussian distribution with zero mean and standard deviation $\sigma_P$.\\
The astrometric problem to be solved now to get the parameters $\vec{\beta}_{a}$ is analogous to Eq. (\ref{eq:hdix})
\begin{equation}
X\vec{\beta}_{a}+\vec{\epsilon}=\vec{w}+\vec{\Delta w}.
\end{equation}
The estimate of the parameters is again found by $\chi^2$ minimization
\begin{equation}
\chi_{a}^2=W^{-1}[(\vec{w}+\vec{\Delta w})-X\vec{\beta}_{a}]^T[(\vec{w}+\vec{\Delta w})-X\vec{\beta}_{a}]
\label{eq:chiasy}
\end{equation}
and the result, in analogy with Eq.(\ref{eq:betaest}), is
\begin{equation}
\vec{\hat{\beta}}_{a}=(X^TW^{-1}X)^{-1}(X^TW^{-1}(\vec{w}+\vec{\Delta w})).
\label{eq:betaasy}
\end{equation}
With the use of  Eqs. (\ref{eq:betaest}) and (\ref{eq:betaasy}), we can produce an expression for the effect of the surface brightness asymmetries on the parameter estimates
\begin{equation}
\vec{\hat{\beta}}_{a}-\vec{\hat{\beta}_{na}}=(X^TW^{-1}X)^{-1}(X^TW^{-1}\vec{\Delta w})=Z(X^TW^{-1}\vec{\Delta w}).
\label{eq:deltabeta}
\end{equation}
For a non-variable star, the elements of $Z$ and $\sigma_{w_i}$, will be approximately all the same as they are a function of the apparent magnitude $G$ and we can rewrite Eqs.~(\ref{eq:betaest}), (\ref{eq:covmat}), and (\ref{eq:deltabeta}) for the derived parameters and the covariance matrix, as
\begin{equation}
\vec{\hat{\beta}_{na}}=(X^TX)^{-1}(X^T\vec{w}),
\end{equation}
\begin{equation}
\vec{\hat{\beta}}_{a}=(X^TX)^{-1}(X^T(\vec{w}+\vec{\Delta w})),
\end{equation}
\begin{equation}
Z=\sigma_{w}^2(X^TX)^{-1}.
\end{equation}
We note that the covariance matrix $Z$ is \textit{independent of the presence of the brightness asymmetries}, which means that the uncertainty in the deduced parameters will be the same. That is because what enters the covariance matrix are basically the measurement errors.\\
In reality, the \textit{accuracy} (to be distinguished from the \textit{precision}) of the parameters can be quite different in the two cases. Indeed, the parameters derived in presence of brightness asymmetries will be different from the ones found for a uniform-disk star, their difference being given by:
\begin{equation}
\vec{\Delta\hat{\beta}}=\vec{\hat{\beta}}_{a}-\vec{\hat{\beta}_{na}}=Z(X^T W^{-1}\vec{\Delta w})=
\frac{1}{\sigma_{w}^2}\frac{1}{d}
\left(
\begin{array}{c}
\sigma^2_{\beta_1}\sum_{i=1}^N X_{i1}D_i\\
\vdots\\
\sigma^2_{\beta_5}\sum_{i=1}^N X_{i5}D_i\\
\end{array}
\right),
\label{ed:deltabeta}
\end{equation}
where $\sigma_{\beta_j}^2$ are the diagonal elements of the covariance matrix $Z$, $X$ is the matrix of the known coefficients described in Eq.~(\ref{eq:Xdef}), and $D$ describes the photocentre motion (see Eq.~\ref{eq:defdeltaw}). In this equation, all the terms are known apart from $\sum_{i=1}^N X_{ij}D_i$. These terms are on average zero because the factors are all uncorrelated and $\langle D_i\rangle=0$. 
As we expect to see some effects on the deduced parameters because of the presence of surface brightness asymmetries, it is interesting to compute $\sigma^2_{\vec{\Delta\hat{\beta}}}$, the standard deviation of $\vec{\hat{\beta_{a}}}-\vec{\hat{\beta}_{na}}$. This quantity gives an estimate of the errors in the value of the derived parameters caused by the presence of the brightness asymmetries. For the parameter $\beta_j$, this means that we need to estimate the variance in $N\langle X_{ij}D_i \rangle_{_i}$.
We can easily compute those factors for all the parameters where  $\langle a \rangle=\langle b \rangle=0$, $\sigma^2_a=\sigma^2_b=\sigma_P^2$ and $\langle t \rangle=0$.\\
For the first two parameters $\alpha_0^*$ and  $\delta$, we have to compute the variance of $\langle X_{ij}D_i \rangle_{_i}= (\sin \theta) D$ and $\langle X_{ij}D_i \rangle_{_i}= (\cos \theta) D$,  which leads to
\begin{equation}
\sigma_{\Delta \hat{\alpha}^*_{0}}=\sigma_{\hat{\alpha}^*_{0-a}-\hat{\alpha}_{0-na}^*}=\sqrt{\frac{N}{2}}\frac{\sigma^2_{\alpha_0^*}}{\sigma_{w}^2}\frac{\sigma_P}{d}
\end{equation}
and
\begin{equation}
\sigma_{\Delta \hat{\delta}_{0}}=\sigma_{\hat{\delta}_{0-a}-\hat{\delta}_{0-na}}=\sqrt{\frac{N}{2}}\frac{\sigma^2_{\delta_0}}{\sigma_{w}^2}\frac{\sigma_P}{d}.
\end{equation}
For the parallax $\langle X_{ij}D_i \rangle_{_i}= f_wD$, we need the variance in the parallax factors, which is easily obtained by noting that the only term that is not constant in time is $\odot$ and neglecting the variations in $r$:
\begin{equation}
\sigma^2_{P_{\alpha}}=\frac{1}{2}(\cos^2\epsilon \cos^2\alpha+\sin^2\alpha),
\label{eq:sigmaPa}
\end{equation}
\begin{equation}
\sigma^2_{P_{\delta}}=\frac{1}{2}\left[(\sin\epsilon\cos\delta-\cos\epsilon\sin\alpha\sin\delta)^2+\cos^2\alpha\sin^2\delta\right].
\label{eq:sigmaPd}
\end{equation}
Hence
\begin{equation}
\sigma_{\Delta\hat{\varpi}}=\sigma_{\hat{\varpi}_{a}-\hat{\varpi}_{na}}=\sqrt{\sigma^2_{P_{\alpha}}+\sigma^2_{P_{\delta}}}\sqrt{\frac{N}{2}}\frac{\sigma^2_{\varpi}}{\sigma_{w}^2}\frac{\sigma_P}{d}.
\end{equation}
Finally for the two components of the proper motion  $\mu_{\alpha^*}$ and  $\mu_{\delta}$, we have to compute the variance in $\langle X_{ij}D_i \rangle_{_i}=t (\sin \theta) D$ and $\langle X_{ij}D_i \rangle_{_i}=t (\cos \theta) D$, which leads to
\begin{equation}
\sigma_{\Delta\hat{\mu}_{\alpha^*}}=\sigma_{\hat{\mu}_{\alpha^*-a}-\hat{\mu}_{\alpha^*-na}}=\sqrt{\sigma^2_t}\sqrt{\frac{N}{2}}\frac{\sigma^2_{\mu_{\alpha^*}}}{\sigma_{w}^2}\frac{\sigma_P}{d}
\end{equation}
and
\begin{equation}
\sigma_{\Delta\hat{\mu}_{\delta}}=\sigma_{\hat{\mu}_{\delta-a}-\hat{\mu}_{\delta-na}}=\sqrt{\sigma^2_t}\sqrt{\frac{N}{2}}\frac{\sigma^2_{\mu_{\delta}}}{\sigma_{w}^2}\frac{\sigma_P}{d}.
\end{equation}
\section{Expression for $\chi^2$ in the presence of surface brightness asymmetries}
We estimate the increase in the $\chi^2$ caused by the presence of surface brightness asymmetries. We can rewrite Eq.~(\ref{eq:chiasy}) for $\chi^2_{a}$ in a slightly different form
\begin{equation}
\chi_{a}^2=W^{-1}[\vec{w}-X\vec{\beta_{na}}+\vec{\Delta w}-X\vec{\Delta \beta})]^T[\vec{w}-X\vec{\beta_{na}}+\vec{\Delta w}-X\vec{\Delta \beta})]
\end{equation}
where $\vec{\Delta \beta}$ is $\vec{\beta}_{a}-\vec{\beta}_{na}$. The minimum $\chi_{a}^2$ will be, by definition, the one for the deduced parameters, hence with $\vec{\Delta \beta}=\vec{\Delta \hat{\beta}}$ (Eq.~(\ref{ed:deltabeta})) and with $\vec{\beta}_{na}=\vec{\hat{\beta}_{na}}$ (Eq.~(\ref{eq:betaest})). Using those values, we obtain
$$
\chi_{a}^2=W^{-1}\left[\left(\vec{w}-X\vec{\hat{\beta}_{na}}\right)^T\left(\vec{w}-X\vec{\hat{\beta}_{na}}\right)+\vec{\Delta w}^T\vec{\Delta w}+\left(X\vec{\Delta \hat{\beta}}\right)^T\left(X\vec{\Delta \hat{\beta}}\right)+\right.
$$
\begin{equation}
\left.-2\vec{\Delta w}^T\left(X\vec{\Delta \hat{\beta}}\right)+2\left(\vec{w}-X\vec{\hat{\beta}_{na}}\right)^T\vec{\Delta w}-2\left(\vec{w}-X\vec{\hat{\beta}_{na}}\right)^T\left(X\vec{\Delta \hat{\beta}}\right)\right].
\label{eq:darisolvere}
\end{equation}
The last four terms are negligible with respect to the first two, leading to
\begin{equation}
\chi_{a}^2\simeq\chi_{na}^2+N\frac{1}{\sigma_w^2}\left(\frac{\sigma_P}{d}\right)^2.
\label{eq:chiasyfin}
\end{equation}

\end{appendix}

\end{document}